\documentclass[iop]{emulateapj} 
\usepackage{apjfonts}

\begin{document}

\epsscale{1.1}

\title{The Discovery of Pulsating Hot Subdwarfs in NGC~2808\altaffilmark{1}}

\shorttitle{Pulsating Subdwarfs in NGC~2808}

\author{
Thomas M. Brown\altaffilmark{2}, 
Wayne B. Landsman\altaffilmark{3}, 
Suzanna K. Randall\altaffilmark{4},
Allen V. Sweigart\altaffilmark{5}, 
Thierry Lanz\altaffilmark{6}
}

\altaffiltext{1}{Based on observations made with the NASA/ESA {\it Hubble
Space Telescope}, obtained at STScI, which
is operated by AURA, Inc., under NASA contract NAS 5-26555.}

\altaffiltext{2}{Space Telescope Science Institute, 3700 San Martin Drive,
Baltimore, MD 21218, USA; tbrown@stsci.edu}

\altaffiltext{3}{Adnet Systems, NASA Goddard Space Flight Center, Greenbelt,
MD 20771, USA; Wayne.Landsman@nasa.gov}

\altaffiltext{4}{European Southern Observatory, 
Karl-Schwarzschild-Str.\ 2, 85748 Garching bei M\"{u}nchen, Germany;
srandall@eso.org}

\altaffiltext{5}{NASA Goddard Space Flight Center, Greenbelt, MD 20771, USA;
allen.sweigart@gmail.com}

\altaffiltext{6}{Laboratoire Lagrange, UMR7293, Universit\'{e} de Nice 
Sophia-Antipolis, CNRS, Observatoire de la C\^{o}te d’Azur, F-06304 Nice, 
France; Thierry.Lanz@oca.eu}

\submitted{Accepted for publication in The Astrophysical Journal Letters}

\begin{abstract}

We present the results of a {\it Hubble Space Telescope} program to
search for pulsating hot subdwarfs in the core of NGC~2808.  These
observations were motivated by the recent discovery of such stars in
the outskirts of $\omega$ Cen.  Both NGC~2808 and $\omega$ Cen are
massive globular clusters exhibiting complex stellar populations and
large numbers of extreme horizontal branch stars.  Our far-UV
photometric monitoring of over 100 hot evolved stars has revealed six
pulsating subdwarfs with periods ranging from 85 to 149~s and UV
amplitudes of 2.0 to 6.8\%.  In the UV color-magnitude diagram of
NGC~2808, all six of these stars lie immediately below the canonical
horizontal branch, a region populated by the subluminous ``blue-hook''
stars.  For three of these six pulsators, we also have low-resolution
far-UV spectroscopy that is sufficient to broadly constrain their
atmospheric abundances and effective temperatures.  Curiously, and in
contrast to the $\omega$ Cen pulsators, the NGC~2808 pulsators do not
exhibit the spectroscopic or photometric uniformity one might expect
from a well-defined instability strip, although they all fall within a
narrow band (0.2~mag) of far-UV luminosity.

\end{abstract}


\section{Introduction}

\subsection{Formation of Extreme Horizontal Branch Stars}

The formation of extreme horizontal branch (EHB) stars has been an
intriguing puzzle for decades (for a review, see Heber 2009).  
These stars are distinguished by their
high effective temperatures ($T_{\rm eff} > $20,000~K) and surface
gravities (log~$g > $5), and are located at the hot end of the HB in
globular clusters (GCs) with extended blue HB morphologies.  Their
analogs in the field, the subdwarf B (sdB) stars, are responsible for
the ``UV upturn'' in the otherwise cool spectra of ellipticals (e.g.,
Brown et al.\ 1997; Yi et al.\ 1997; O'Connell 1999; Brown et
al.\ 2000).  EHB stars have extremely thin envelope masses ($<
10^{-2}$~$M_\odot$), implying that they have undergone significant
mass loss on the red-giant branch (RGB) since leaving the main
sequence (MS).  The challenge has been to understand how a star can
lose $\sim$0.3~$M_\odot$ on the RGB while retaining enough mass to
ignite helium in the core and evolve to the EHB.  However, observations
to date have not determined how this mass loss actually occurs.  The
high fraction of binaries among field subdwarfs (e.g., Maxted et al.\ 2001)
suggests a binary mechanism for their production such as 
Roche Lobe overflow,
common envelope evolution, white dwarf (WD) mergers (e.g., Han et
al.\ 2002, 2003), or MS-WD mergers (e.g., Clausen \& Wade 2011). 

New insight into the formation of EHB stars in GCs has come from the
discoveries of a double MS in $\omega$ Cen (Anderson
1997) and a triple MS in NGC~2808 (D'Antona et al.\ 2005; Piotto et
al.\ 2007).  These massive GCs apparently contain a significant
($\sim$20\%) population of helium-rich ($Y \sim 0.4$) stars (Piotto et
al.\ 2005; Dupree \& Avrett 2013), likely formed in a second stellar
generation from the helium-rich ejecta of the first generation (see
Bekki \& Norris 2006). With enough scrutiny, one can find at least
small variations in chemical composition for any GC, but the massive
GCs exhibit significantly larger spreads in many heavy elements as
well as clear evidence for helium-rich subpopulations (for a review,
see Gratton et al.\ 2012).  These helium-rich subpopulations may
explain why massive GCs have HB morphologies that extend to high
$T_{\rm eff}$ (D'Antona et al.\ 2002; Busso et al.\ 2007).  Because
the MS turnoff mass decreases strongly with increasing helium for a
given GC age, a helium-rich star will arrive on the HB with a
significantly lower mass and therefore a higher $T_{\rm eff}$ than a
helium-normal star, assuming the same RGB mass loss.  Thus, single-star
mass loss in the helium-rich stars via a strong stellar wind
(e.g., Schr\"{o}der
\& Cuntz 2005), or even planet ingestion (Soker 1998), may
be sufficient to populate the EHB.  This possibility is supported by the 
low binary fraction among the GC subdwarfs 
(Moni Bidin et al.\ 2006, 2009, 2011).  Moreover,
the fact that the fraction of HB stars falling on the EHB does not appear
to vary significantly with radius in $\omega$ Cen or NGC~2808 (e.g.,
Whitney et al.\ 1998; Iannicola et al.\ 2009) 
would also argue against a binary
origin for these EHB stars.  It appears therefore that the EHB stars in 
massive GCs are most likely the progeny of the most helium rich subpopulations,
in contrast to the field population, where a greater diversity of formation
mechanisms may be at play.

\begin{figure}[t!]
\plotone{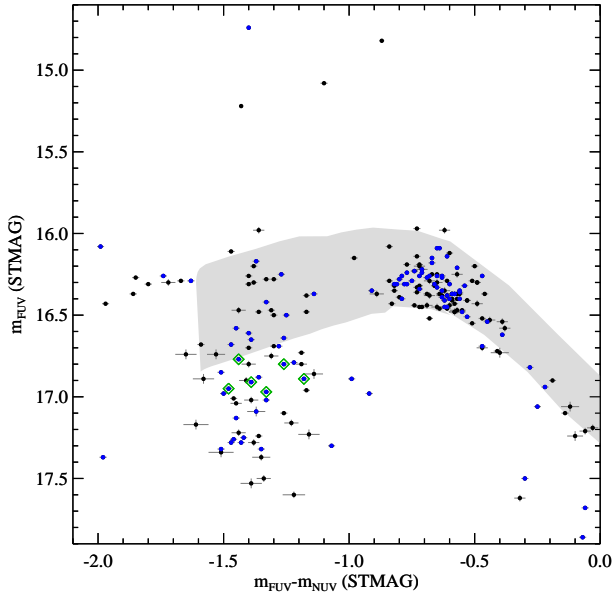}
\caption{The UV CMD of NGC~2808, in the STIS FUV/F25QTZ and NUV/F25CN270
  bands.  The full sample of Brown et al.\ (2001) is shown by blue and black
  points, with photometric error bars indicated.  
  The canonical HB locus
  is indicated by grey shading, bounded at the bottom 
  by the zero-age HB (ZAHB) and at the top by
  the point at which stars have completed 99\% of the core
  helium-burning lifetime.  The bluest point on the canonical ZAHB,
  at $m_{\rm FUV} - m_{\rm NUV} = -1.59$~mag, corresponds to a $T_{\rm eff}$
  of 31,615~K, but the correspondence between UV color and $T_{\rm eff}$
  for any given star depends heavily upon the chemical composition of the 
  atmosphere. NGC~2808 has a significant population of
  blue-hook stars lying below the hot end of the canonical HB; although
  the UV colors of the blue-hook stars are similar to the colors on the
  canonical EHB, the blue-hook stars can be much hotter.  Blue points indicate
  those stars monitored in our time-tag photometry.  The six
  pulsators, highlighted with green diamonds, span a wide range of
  color in the blue-hook region of the UV CMD.}
\end{figure}

Recent observations have revealed that massive GCs also host a
population of subluminous EHB stars that cannot be explained by the
usual post-RGB evolution.  These subluminous stars were first
discovered in $\omega$~Cen, where they form a ``blue hook'' at the hot
end of the EHB (D'Cruz et al.\ 2000).  D'Cruz et al. (2000) proposed
that these blue-hook stars were the result of a delayed helium-core
flash beyond the RGB tip.  Brown et al.\ (2001) subsequently found a
large population of blue-hook stars in the UV color-magnitude diagram
(CMD) of NGC~2808 (Figure 1).  Using new evolutionary and
spectroscopic models, Brown et al.\ (2001) demonstrated that these
blue-hook stars were the result of flash mixing on the WD cooling
curve.  Normally, stars ignite helium at the RGB tip (Figure 2), but
stars that lose sufficient mass, either through single- or binary-star
mechanisms, will leave the RGB and evolve to high $T_{\rm eff}$ before
igniting helium (Castellani \& Castellani 1993).  If the helium-core
flash is delayed until the WD cooling curve, the flash convection will
mix the hydrogen envelope into the helium core (Sweigart 1997),
leading to greatly enhanced surface abundances of helium and carbon as
well as a much higher $T_{\rm eff}$ during the subsequent EHB phase.
The higher $T_{\rm eff}$, together with the decreased hydrogen opacity,
make the flash-mixed stars subluminous in observing bands longward of the 
Lyman limit, when compared to normal HB stars.

In a later study, Brown et al.\ (2010) found blue-hook stars in 5 other
massive GCs spanning a wide range of metallicity.  Spectroscopic
investigations of the blue-hook and normal EHB populations in
$\omega$~Cen (Moehler et al.\ 2011) and NGC~2808 (Brown et al.\ 2012)
have confirmed that, compared to the normal EHB stars, the blue-hook
stars are both much hotter and enhanced in helium and carbon, thus
providing unambiguous evidence for flash mixing.  From an analysis of
Space Telescope Imaging Spectrograph (STIS) spectra, Brown et
al.\ (2012) found that the hottest blue-hook stars in NGC~2808 exhibit
effective temperatures up to 50,000~K, carbon abundances orders of
magnitude higher than in normal EHB stars, helium abundances of ~99\%
by mass, and enormous enhancements in the iron-peak elements from
radiative levitation.

\begin{figure*}[t!]
\plotone{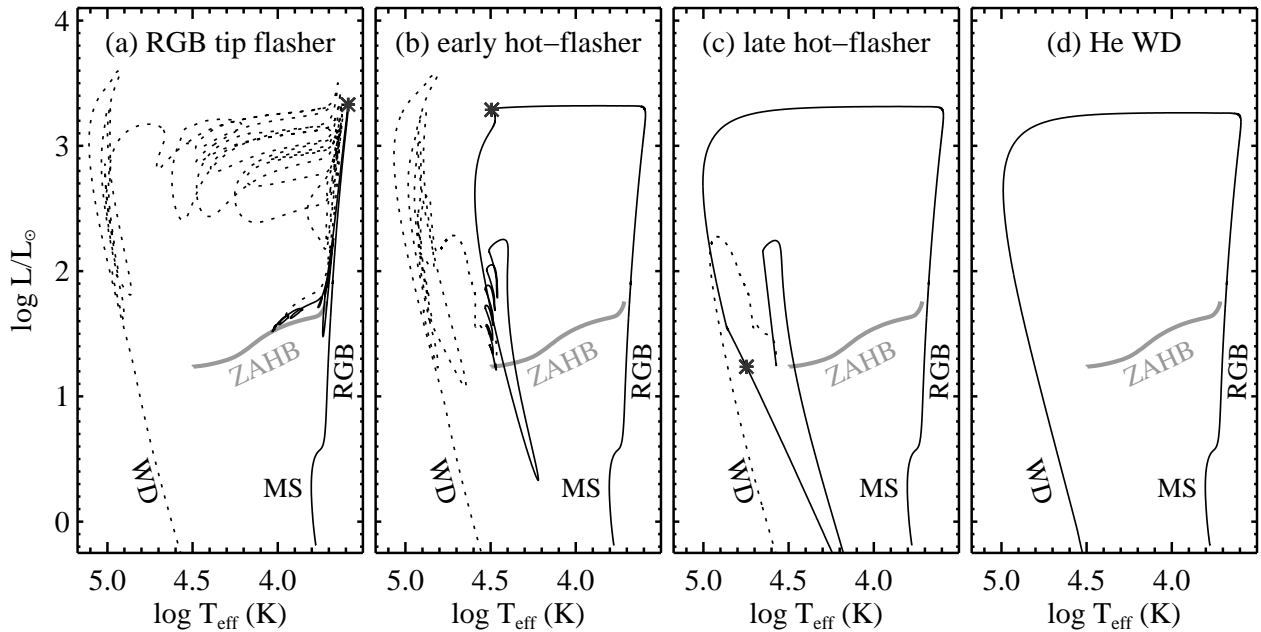}
\caption{Various evolutionary paths for producing an HB star 
  (from Brown et al.\ 2010). The ZAHB phase is highlighted in
  grey, while the pre-ZAHB evolution (solid curves) and post-ZAHB
  evolution (dashed curves) are in black. The peak of the helium-core
  flash is marked by an asterisk. The four panels show the evolution
  for progressively larger amounts of mass loss on the RGB. The
  evolution in the first two panels produces canonical HB stars in
  which the hydrogen-rich surface composition does not change during the
  helium-core flash. In the third panel, the helium-core flash occurs
  on the WD cooling curve, producing a flash-mixed star having a
  surface composition highly enriched in helium and carbon and a
  temperature significantly hotter than the canonical HB.  In the
  fourth panel, the helium flash never occurs and the star dies as a
  helium WD.}
\end{figure*}

\subsection{Pulsating Subdwarfs}

One of the exciting new avenues for exploring the formation and
evolutionary status of hot subdwarfs has been asteroseismology.
Rapidly-pulsating hot subdwarfs are well-established in the field, where they
are known as V361 Hya stars (also known as EC 14026 stars; 
Kilkenny et al.\ 1997).  Their short ($\sim$100--200~s)
non-radial pulsations are driven by an iron opacity mechanism
associated with a local overabundance of iron in the driving region
(Charpinet et al.\ 1996), 
produced by diffusion in these otherwise metal-poor stars.
The V361 Hya stars inhabit an instability strip at 29,000~K $\lesssim
T_{\rm eff} \lesssim $ 36,000~K (Charpinet et al.\ 1997).  Asteroseismic
analysis of hot subdwarfs can yield accurate structural parameters
(e.g., $T_{\rm eff}$ to 0.6\%, log~$g$ to 0.03\%, total mass to
  1\%, and envelope mass to 20\%; Charpinet et al.\ 2009).  For this reason,
the Kepler mission has dedicated significant observing time toward
asteroseismology of hot subdwarfs in the field
(e.g., {\O}stensen et al.\ 2010).

Asteroseismology in GCs has been a different story.  Until very
recently, all searches for pulsating subdwarfs in GCs had been unsuccessful
(e.g., Reed et al.\ 2006).  However, Randall et al.\ (2011, 2012) have
now discovered 5 hot pulsating subdwarfs in $\omega$~Cen.  Compared to
the field pulsators, these new pulsators have somewhat shorter periods
(84--124~s) and higher temperatures ($\sim$50,000~K), with similar pulsation
amplitudes ranging from 0.9 to 2.7\%.  The $\omega$ Cen pulsators
raise the exciting possibility of a new instability strip unseen among
the field subdwarfs.  Given its size, distance, and reddening, NGC~2808 is
the most logical candidate to search for such pulsators beyond
$\omega$ Cen.  Previous far-UV photometry and spectroscopy of the
NGC~2808 core revealed a large population of hot evolved stars to
check for pulsations, including blue-hook stars near the temperature
of the $\omega$ Cen pulsators (Brown et al.\ 2001, 2012).  In this
Letter, we report the discovery of six new pulsators among the
hot evolved stars of NGC~2808.  

\begin{figure}[b!]
\plotone{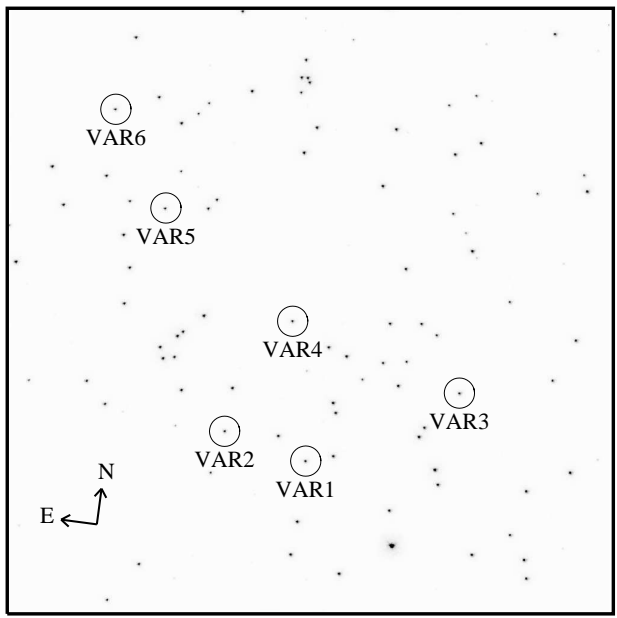}
\caption{
The STIS FUV image of the NGC~2808 core, created from a sum of the entire
series of time-tag data, with a linear stretch.  Short-period
pulsators are indicated by circles and labeled.  Although this is the
core of a massive globular cluster, the suppression of the dominant
cool population by the F25SRF2 filter results in a sparse field where
accurate time-series photometry is possible.}
\end{figure}

\section{Observations and Data Reduction}

We obtained far-UV images of the NGC~2808 core in 5 consecutive orbits
on 28 March 2013, using STIS on the {\it Hubble Space Telescope
  (HST)}.  The cumulative exposure time was 16 ks, spanning a 26
ks observing window interrupted by four 2.5 ks gaps due to
occultations.  Although we used the same instrument employed by Brown
et al.\ (2001), the new far-UV imaging of the NGC~2808 core was
distinct in three ways.  First, only a single
25$\times$25$^{\prime\prime}$ field was imaged within the larger area
imaged previously.  Second, we used the time-tag mode on STIS, which
records the arrival time of each detected photon to an accuracy of 125
$\mu$s.  Third, we used the FUV/F25SRF2 bandpass instead of the
FUV/F25QTZ bandpass.  Because the F25SRF2 bandpass extends to
shorter wavelengths, it includes the terrestrial \ion{O}{1}
$\lambda$1301 emission, increasing the background from $\sim$10$^{-6}$
counts s$^{-1}$ pixel$^{-1}$ to $\sim$10$^{-3}$ counts s$^{-1}$
pixel$^{-1}$.  Although the F25SRF2 background is still quite low, it
varies with orbital day and night, and is significant when integrated
over the entire detector, making management of the STIS data buffer in
time-tag mode more difficult.  The advantage is that the throughput in
the F25SRF2 bandpass is much higher than that in the F25QTZ bandpass.
For reference, a star with $T_{\rm eff}$~=~30,000~K at
$m_{\rm FUV}$~=17~mag in the F25QTZ bandpass (e.g., see Figure 1) would
produce 4.5 counts s$^{-1}$ in the F25QTZ bandpass and 18 counts
s$^{-1}$ in the F25SRF2 bandpass.  Because we monitored 110 hot stars
that are bright in the UV, the photometric errors are dominated
by Poisson noise on the source counts, and so the use of the broader
bandpass enabled the detection of significantly weaker pulsations.
In general, 
the use of a far-UV bandpass suppresses the dominant cool population,
enabling accurate photometry in a sparse field of hot stars (Figure 3),
even in the cluster core.

The STIS data were processed through the standard calibration
pipeline, including corrections to the photon arrival times for
general relativistic effects, displacement of {\it HST} from Earth
center, and displacement of Earth from the solar system barycenter.
The net count rate for each star was measured in two-second bins using
aperture photometry and a local sky subtraction.  We then performed a
Lomb-Scargle Normalized Periodogram on the time-series photometry for each
star, searching for periodic signals with $>$99\% significance and
periods between 50 and 300~s.  This procedure produced six pulsators (Figure
1) for additional analysis.  Three of these pulsators were
included in the spectroscopic sample of Brown et al.\ (2012),
who determined $T_{\rm eff}$ to an accuracy of $<$5,000~K.  In addition,
all of these pulsators have both far-UV
and near-UV photometry from Brown et al.\ (2001).  Strictly speaking, 
the UV photometry provides no constraints on surface gravity and only a lower
limit ($>$20,000~K) on $T_{\rm eff}$, due to the abundance variations in 
this population. However, given the UV colors, VAR2 might be as
hot as $\sim$50,000~K, while VAR3 and VAR4 might be as hot as $\sim$40,000~K.

The light curves of each pulsator were converted to fractional
intensity variations about zero and their Fourier spectra (Figure 4)
were computed in the frequency range of interest (0 -- 15 mHz).  Any
peaks above a detection limit set to four times the local noise level
were successively pre-whitened. For this, the measured period was kept
fixed, while the amplitude and phase were determined using a
least-squares fitting procedure to the light curve. The resulting
sinusoid was subtracted from the light curve, the Fourier spectrum was
re-computed, and the exercise was repeated until no peaks remained
above the detection threshold. Using this method, we extracted the
periods and amplitudes listed in Table 1.  The dominant period for
each pulsator is the same as the one detected 
in the initial Lomb-Scargle
Periodogram.  The amplitude uncertainty is 0.4\%, corresponding to the
formal least-squares fitting errors.  The period uncertainty is
$\sim$0.1~s, corresponding to a tenth of the frequency resolution of
the data set.  Besides the periodicities characterized in Table 1, the
Fourier spectra (Figure 4) exhibit additional peaks above the
detection threshold, but these are aliases of the true peaks, offset
by 0.17 mHz (corresponding to the time separation between the
contiguous observing blocks).  While the Lomb-Scargle periodogram of VAR5
implied a significant signal of 146.8~s, the detection in the
Fourier analysis is marginal (2.0\% amplitude), being just below the
4$\sigma$ threshold (2.1\% amplitude).  Splitting the VAR5 time series into
two halves, the 146.8~s periodicity appears in the independent 
Fourier spectra of each half, lending further credence to the signal.
We also note here (but not in
Table 1) a tentative detection of a second periodic signal for VAR2,
with a period of 129.5~s and an amplitude of 1.8\%.

\begin{figure}[t!]
\plotone{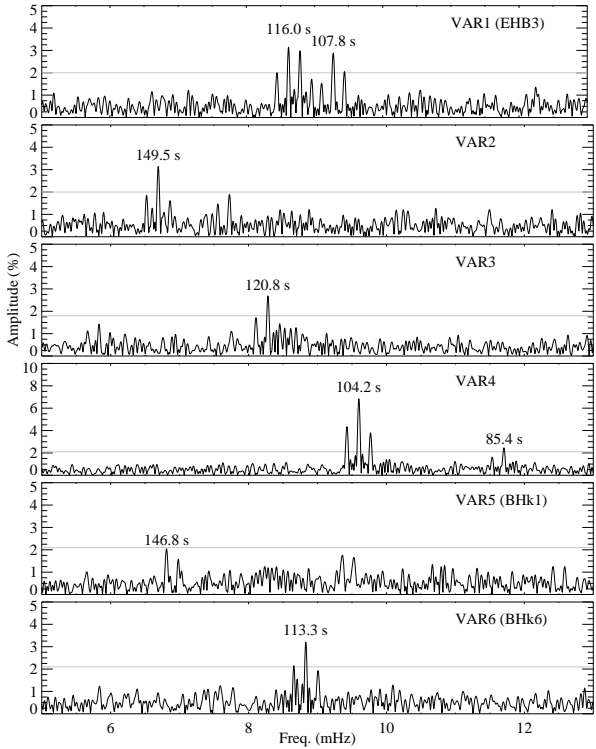}
\caption{Fourier amplitude spectra of the 6 pulsators in NGC~2808,
  with the significant periodic signals labeled for each.  Other peaks
  are aliases of the true periodic signals.  Three of the six pulsators
  have far-UV spectroscopy from Brown et al.\ (2012), with the names
  from that study indicated (EHB3, BHk1, BHk6).  The 4$\sigma$
  detection limit in the Fourier analysis is indicated by a grey line.}
\end{figure}

\begin{table*}[t!]
\caption{Properties of NGC~2808 Pulsators}
\hskip 0.9in
\begin{tabular}{llcccccc}
\tableline
     & Alternate & UV Catalog & Periods & Amplitudes  & $T_{\rm eff}$\tablenotemark{a} &        & Fe-peak \\
Name & Name\tablenotemark{a} & Entry\tablenotemark{b}      & (s)     & (\%)       & (K)          & $Y$\tablenotemark{a}   & enhancement\tablenotemark{a} \\
\tableline
VAR1& EHB3& 149& 116.0, 107.8& 3.1, 2.9 &  30,000 & 0.23& 10$\times$ \\
VAR2& --  & 185& 149.5       & 3.1      & $\sim$20,000--50,000 & --  & --         \\
VAR3& --  &  85& 120.8       & 2.7      & $\sim$20,000--40,000 & --  & --         \\
VAR4& --  & 164& 104.2, 85.4 & 6.8, 2.3 & $\sim$20,000--40,000 & --  & --         \\
VAR5& BHk1& 230& 146.8       & 2.0      &  50,000 & 0.99& 50$\times$ \\
VAR6& BHk6& 250& 113.3       & 3.2      &  25,000 & 0.23& 10$\times$ \\ 
\tableline
\end{tabular}

\hskip 0.9in $^{a}$Brown et al.\ (2012)

\hskip 0.9in $^{b}$Brown et al.\ (2001)
\end{table*}

Our observing program was designed to detect pulsators on the EHB 
(at $>$99\% significance) if they have periods of order 100~s and
amplitudes of at least 2\%.  Thus, we cannot exclude the possibility
that pulsators with lower amplitudes might exist.  While our pulsators
are all in the blue-hook region of the UV CMD (Figure 1), and are thus
somewhat fainter than the canonical EHB, our completeness in detecting
pulsations is not a
strong function of UV luminosity, and is similar for both the EHB and
blue-hook stars.  To quantify the completeness, we shuffled the
observed light curves for our six pulsators (while maintaining their
occultations), inserted a periodic signal, and tried to blindly
recover that signal using the Lomb-Scargle periodogram screening described
above.  For the six pulsators, amplitudes in the range of 1.8\% to
2.0\% resulted in significant ($>$99\%) detections in half the trials,
with slight star-to-star variations.  For larger amplitudes, the
sample quickly becomes more complete, with $>$95\% completeness for
amplitudes above 2.5\%.  Artificial signal tests of the time-series
photometry for the other stars in our sample yielded similar results.
For the faintest blue-hook star and brightest EHB star in our sample,
significant detections were made in half the trials for amplitudes of
2.4\% and 1.4\%, respectively.
  
We note that the point spread function (PSF) of VAR4, which is the
pulsator exhibiting the strongest pulsations, is partly shadowed by
the STIS repeller wire.  This wire casts a thin shadow across the STIS
FUV detector, with a depth of $\sim$10\%.  However, that shadowing
cannot account for the variations observed in the photometry of this
star.  The shadow falls approximately 0.1$^{\prime\prime}$ from the
center of the PSF, and the star did not drift significantly relative
to the repeller wire over the entire imaging period.  The region of
the PSF impacted by the repeller wire accounts for 6\% of the
encircled energy of the PSF, but given the depth of the shadow, plus
its stability over the observations, the impact on the time-series
photometry would be $<<$1\%.  We simply note the presence of the shadow
for completeness.

\section{Discussion}

Surprisingly, the six NGC~2808 pulsators form a rather inhomogeneous
group.  Their periods range from 85 to 149~s, with amplitudes of 2.0
to 6.8\%, and span 0.3~mag in UV color.  The 3 pulsators with spectra
span a $T_{\rm eff}$ range of 25,000 to 50,000~K, with both
helium-rich and hydrogen-rich atmospheres present.  VAR5 is only the
second pulsating subdwarf known to be enhanced in helium; the first
was LSIV-14$^{\rm o}$116, a slowly-pulsating (period $>$2000~s)
He-sdB star with $n_{\rm He}$=0.21, which is not a particularly high
helium enhancement for the He-sdB class (Ahmad \& Jeffery 2005).  All
of the NGC~2808 pulsators with spectroscopy 
exhibit the enhancement of the iron-peak elements (relative to
the cluster abundance) that is a common product of radiative diffusion
in hot subdwarfs.

The only obvious commonality among the NGC~2808 pulsators is that they
lie in a narrow range of far-UV luminosity, immediately below the
ZAHB, among the blue-hook stars in the UV CMD.  The luminosity of one
of these stars (VAR1) is close enough to the ZAHB that it was
identified as a normal EHB star by Brown et al.\ (2012).  Neither VAR1
nor VAR6 exhibit an enhanced helium abundance, as would be expected for
stars that have undergone flash mixing.  Perhaps the surface helium abundance
in these stars was depleted through
atmospheric diffusion; alternatively, these stars may not have
undergone flash mixing at all.  The fact that the six pulsators span a
range in $m_{\rm FUV}$ that is only 0.2~mag wide, despite the fact that
the blue-hook stars and normal EHB stars together span an FUV
luminosity range that is nearly ten times larger, may be a clue to the
origin of these stars.  The band of detected pulsators may be bounded
on the bright end by the boundary between normal EHB and blue-hook
stars.  It is not clear what defines the faint luminosity boundary of the
pulsators.  Although pulsators are more difficult to detect at fainter
luminosities, and the amplitude of the pulsations is near our
detection limit, the completeness of pulsation detection is not a
steep function of luminosity across this luminosity range, as
discussed in the previous section.

It appears that the NGC~2808 pulsators are distinct from both the
diverse population of pulsators in the field and the 
homogeneous sample of $\omega$ Cen pulsators (Randall et al.\ 2011),
which are all helium-poor and have $T_{\rm eff}$ near 50,000~K.  The
relatively high yield of pulsating stars in our NGC~2808 survey may be
due to selection bias, given that we focused the search on a sample
consisting solely of hot subdwarfs, and because the
pulsation amplitudes are expected to be stronger in the UV than in the
optical.  Indeed, any distinctions drawn between the NGC~2808
pulsators and the $\omega$ Cen pulsators should be tempered by the
fact that the pulsators were characterized via very distinct data and
atmospheric diagnostics in each cluster; the NGC~2808 pulsators were
found in the cluster core and characterized via UV imaging and
spectroscopy, while the $\omega$~Cen pulsators were found in the
cluster outskirts and characterized via optical imaging and
spectroscopy.  Due to the crowding, optical spectroscopy for the
NGC~2808 pulsators is impossible, even with {\it HST}, but one obvious
avenue for further exploration would be UV spectroscopy of the
$\omega$~Cen pulsators, putting them on the same observational footing
as the NGC~2808 pulsators.  We will also investigate the driving mechanism 
for the NGC~2808 pulsators in our future modeling efforts.  

\acknowledgements 

Support for Program 12954 was provided by NASA through a grant from
STScI, which is operated by AURA, Inc., under NASA contract NAS
5-26555.  The authors are grateful to Marc Rafelski, who graciously
used his software to calculate the orbital variation in the Sun and
target limb angles for the planned observation date, so that we could
confirm the observing window would not suffer from unusually high
\ion{O}{1} emission.

\end{document}